\documentclass{article}

\usepackage{arxiv}

\usepackage[utf8]{inputenc} 
\usepackage[T1]{fontenc}    
\usepackage[pagebackref,breaklinks,colorlinks]{hyperref}
\usepackage{url}            
\usepackage{booktabs}       
\usepackage{amsfonts}       
\usepackage{nicefrac}       
\usepackage{microtype}      
\usepackage{lipsum}

\usepackage{amssymb}
\usepackage{amsmath}
\usepackage[normalem]{ulem}
\usepackage{multirow}
\usepackage{array}
\usepackage{graphicx}
\usepackage{bm}
\usepackage{caption,subcaption}        
\usepackage{pifont}
\usepackage{xcolor}                    

\title{Applications and Challenges of AI and Microscopy in Life Science Research: A Review}

\author{
    Himanshu Buckchash \\
    University of Applied Sciences\\
    Krems, Austria \\
    \texttt{himanshu.buckchash@imc.ac.at} \\
    \And
    Gyanendra Kumar Verma \\
    National Institute of Technology \\
    Raipur, India \\
    \texttt{gkverma.it@nitrr.ac.in} \\
    \And
    Dilip K. Prasad \\
    UiT The Arctic University of Norway \\
    Tromsø, Norway \\
    \texttt{dilip.prasad@uit.no} \\
}

\begin{document}
\maketitle

\begin{abstract}
The complexity of human biology and its intricate systems holds immense potential for advancing human health, disease treatment, and scientific discovery. However, traditional manual methods for studying biological interactions are often constrained by the sheer volume and complexity of biological data. Artificial Intelligence (AI), with its proven ability to analyze vast datasets, offers a transformative approach to addressing these challenges. This paper explores the intersection of AI and microscopy in life sciences, emphasizing their potential applications and associated challenges.
We provide a detailed review of how various biological systems can benefit from AI, highlighting the types of data and labeling requirements unique to this domain. Particular attention is given to microscopy data, exploring the specific AI techniques required to process and interpret this information. By addressing challenges such as data heterogeneity and annotation scarcity, we outline potential solutions and emerging trends in the field.
Written primarily from an AI perspective, this paper aims to serve as a valuable resource for researchers working at the intersection of AI, microscopy, and biology. It summarizes current advancements, key insights, and open problems, fostering an understanding that encourages interdisciplinary collaborations. By offering a comprehensive yet concise synthesis of the field, this paper aspires to catalyze innovation, promote cross-disciplinary engagement, and accelerate the adoption of AI in life science research.
\end{abstract}


\section{Introduction}
Life science is a core research domain with profound implications for human well-being. It addresses myriad questions across diverse directions, all ultimately geared towards advancing human health. However, resolving these challenges is time intensive, as countless experiments are needed to develop effective therapies or robust understanding, each generating extensive data. Manually analyzing such data demands substantial expertise (already in short supply) and is exceedingly time consuming. A notable example is the decades of labor by researchers to empirically determine around 100000 protein structures \cite{alphafold2021}. Recent years, however, have witnessed substantial advances through AI \cite{segmentanything2024medical,alphafold2021}. Utilizing AI to propel life science research is inevitable, given their mutual synergy. AI relies on extensive data to refine its predictive models, while life science generates vast datasets that exceed feasible manual analysis. This synergy is reflected in numerous breakthroughs \cite{alphafold2021,holostain2020,superres2021pssrsimulation,segmentanything2024medical}. This paper aims to elucidate key challenges in microscopy and life science, their interconnections, how AI and microscopy can be harnessed to address them, and prospective research directions.

\begin{figure}[t]
\includegraphics[width=\textwidth]{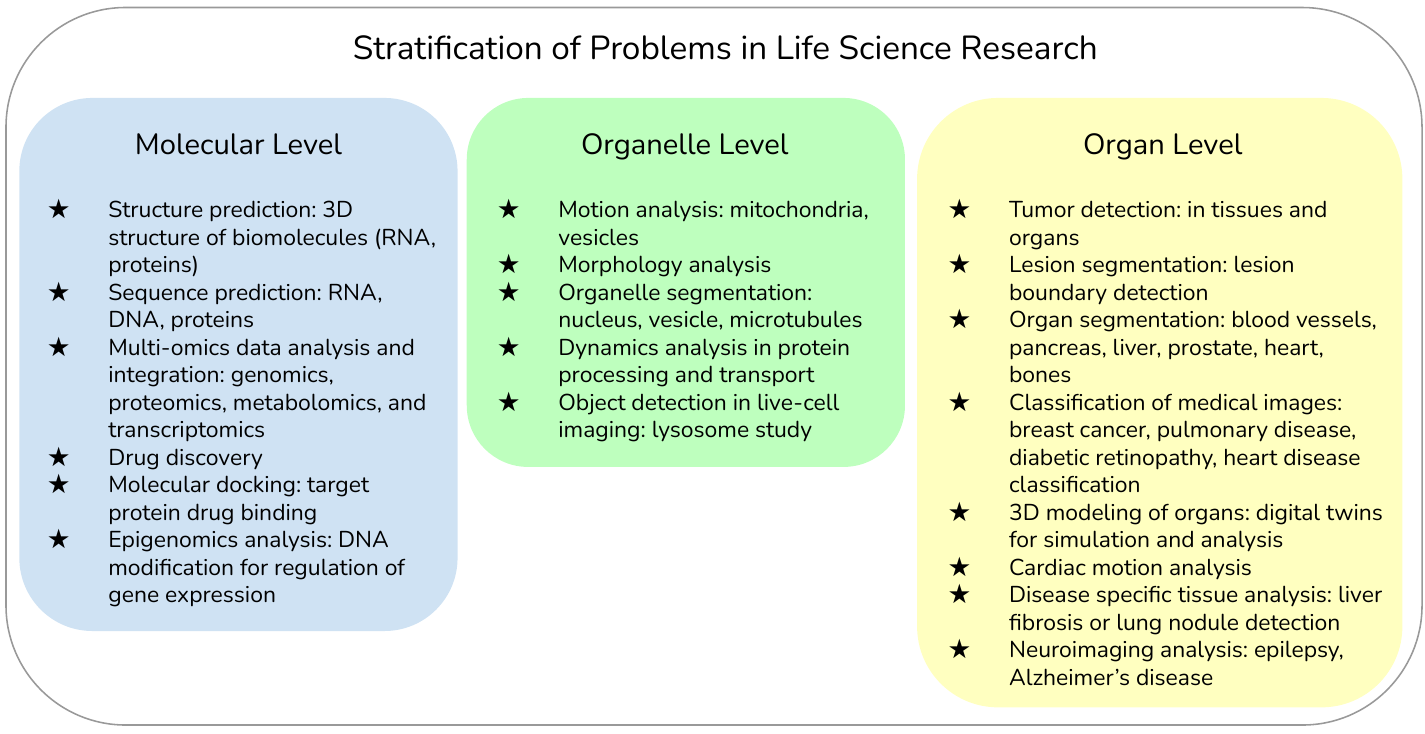}
\caption{A classification of challenges in life sciences, categorized into molecular, organelle, and organ levels. Each tier highlights representative problems.}
\label{fig: 3 tier stratification}
\end{figure}

As illustrated in Fig.~\ref{fig: 3 tier stratification}, the major challenges in life science research (LSR) can be stratified into three hierarchical tiers. The first category encompasses sequence- and structure-focused tasks, such as protein or nucleic acid sequence prediction and omics data analysis, all of which revolve around the molecular level. Algorithms typically employed in these tasks include recurrent neural networks (RNNs), transformers, or graph neural networks, due to their ability to capture complex relationships within molecular data. At a higher spatial scale, the second tier involves subcellular components, including organelles, and aims to elucidate their morphology and dynamics to better understand connections between cellular function and overall well being of the organism. These investigations predominantly rely on microscopic techniques, positioning the study of organelle behavior at the organelle tier, as highlighted in Fig.~\ref{fig: 3 tier stratification}. Finally, the third tier addresses problems at the tissue or organ scale, often employing advanced imaging modalities, such as microscopy and high frequency waves, for visualization. These inquiries typically target organ level health issues, such as tumor detection, and can therefore be classified under the organ level tier. Several examples of these challenges are illustrated in Fig.~\ref{fig: 3 tier stratification}.

In recent years, generative AI has significantly advanced solutions at the molecular level by providing superior outcomes. Models such as \cite{prollamallm2024,hyenadnallm2024} have demonstrated outstanding performance in sequence prediction tasks for proteins, nucleic acids, and related biomolecules. Moreover, graph-based approaches and recurrent neural networks have been employed for modeling multiomics and other complex problems \cite{alphafold2021}, leading to notable progress in disease prognosis, diagnosis, and omics research. However, the integration of AI into microscopy based diagnostics has yet to match these developments, leaving a broad scope of challenges that demand innovative AI driven solutions. Even existing AI models can substantially accelerate microscopy related studies in LSR. This paper aims to explore the intersection of AI and microscopy, emphasizing how these domains can converge to address significant hurdles in LSR. It discusses the key challenges, highlights open research questions, and proposes potential strategies. Written primarily from an AI perspective, this work provides a succinct yet thorough review for researchers in both AI and microscopy, maintaining a balance between technical depth and accessibility. The main objective is to offer a critical perspective for quick engagement with the field. The subsequent sections delve deeper into AI driven imaging, illustrating the synergy between AI and microscopy, scrutinizing key challenges, examining potential remedies, identifying open research directions, describing public resources, and finally presenting concluding remarks.

\section{Synergetic relation between microscopy and AI}
\label{sec:synergetic relation}
\begin{figure}[t]
    \centering
    \begin{subfigure}[b]{0.53\textwidth}
        \centering
        \includegraphics[width=\textwidth]{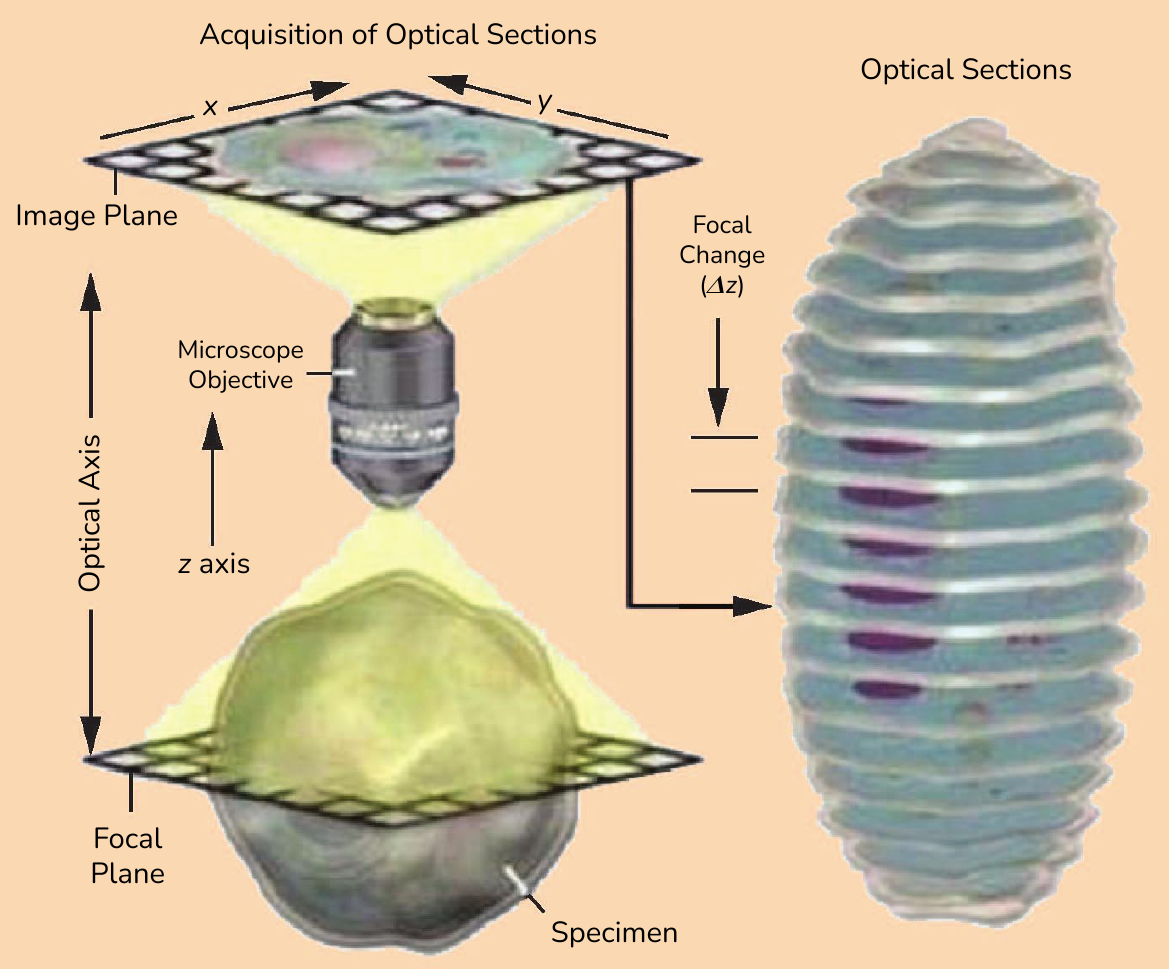}
        \caption{General microscope setup}
        \label{subfig: microscope setup}
    \end{subfigure}
    \hfill
    \begin{subfigure}[b]{0.44\textwidth}
        \centering
        \includegraphics[width=\textwidth]{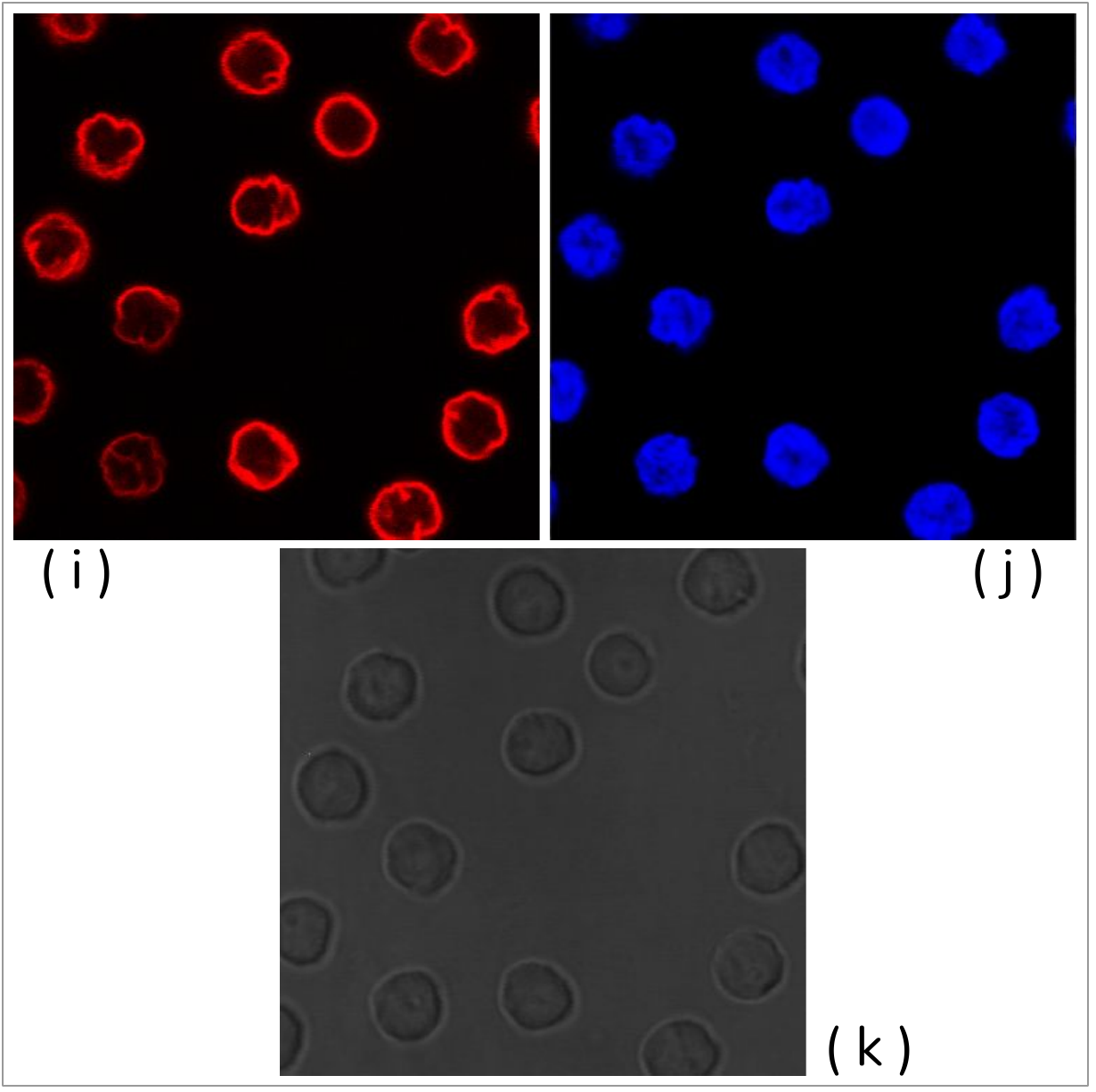}
        \caption{Fluorescence vs Phase-contrast}
        \label{subfig: microscopy image types}
    \end{subfigure}
    \caption{\textbf{Left}: General assembly of a microscope: a schematic representation \cite{microscopegeneralassembly2006}. \textbf{Right}: Microscopic images of culture of human lymphocyte cells. (i) fluorescence image of nuclear envelopes, (j) fluorescence image of interior nuclei (DNA), and (k) phase-contrast image of whole cells. \cite{microscopytypes2022algorithms}.}
    \label{fig:model}
\end{figure}

As depicted in Fig.~\ref{subfig: microscope setup}, a typical microscopy setup involves preparing the specimen on a slide and illuminating it using an appropriate light or wave source. The objective lens then magnifies the resulting signal, and variations in wave patterns or fluorescence provide insights into the sample’s structural and functional features. These interactions enable the formation of detailed images that reveal critical information about the underlying biology. Such images can be exceptionally large and complex, requiring substantial human effort to interpret. Indeed, a single dataset might encompass millions or even billions of subcellular interactions, reflecting intricate biological dynamics \cite{organvisionproject}. Microscopy outputs frequently take the form of 3D, 4D, or 5D tensor data, capturing spatial, temporal, and sometimes spectral dimensions. The sheer scale and complexity of these multidimensional datasets highlight the growing need for AI, fueled both by the success of unsupervised and self-supervised algorithms and the scarcity of labeled data available for training.

AI can play a pivotal role in microscopy by modeling the morphological state and dynamics of subcellular structures, thereby facilitating a deeper understanding of underlying biological processes. Recent efforts \cite{organvisionproject} are dealing with the enormous scale of these interactions to elucidate how diseases manifest in living organisms. A common strategy in such investigations involves applying a \emph{perturbation} --- a controlled alteration or stimulus --- to the organism or system of interest \cite{perturbation2022,nmi2021dkp}. Subsequently, researchers monitor how this perturbation propagates across one or multiple tiers of biological organization, from molecules and organelles to entire organs (see Fig.~\ref{fig: 3 tier stratification}). By capturing the resulting changes at one or multiple levels, this approach provides a more comprehensive perspective on the complex interplay that governs health and disease.

\begin{figure}[t]
\includegraphics[width=\textwidth]{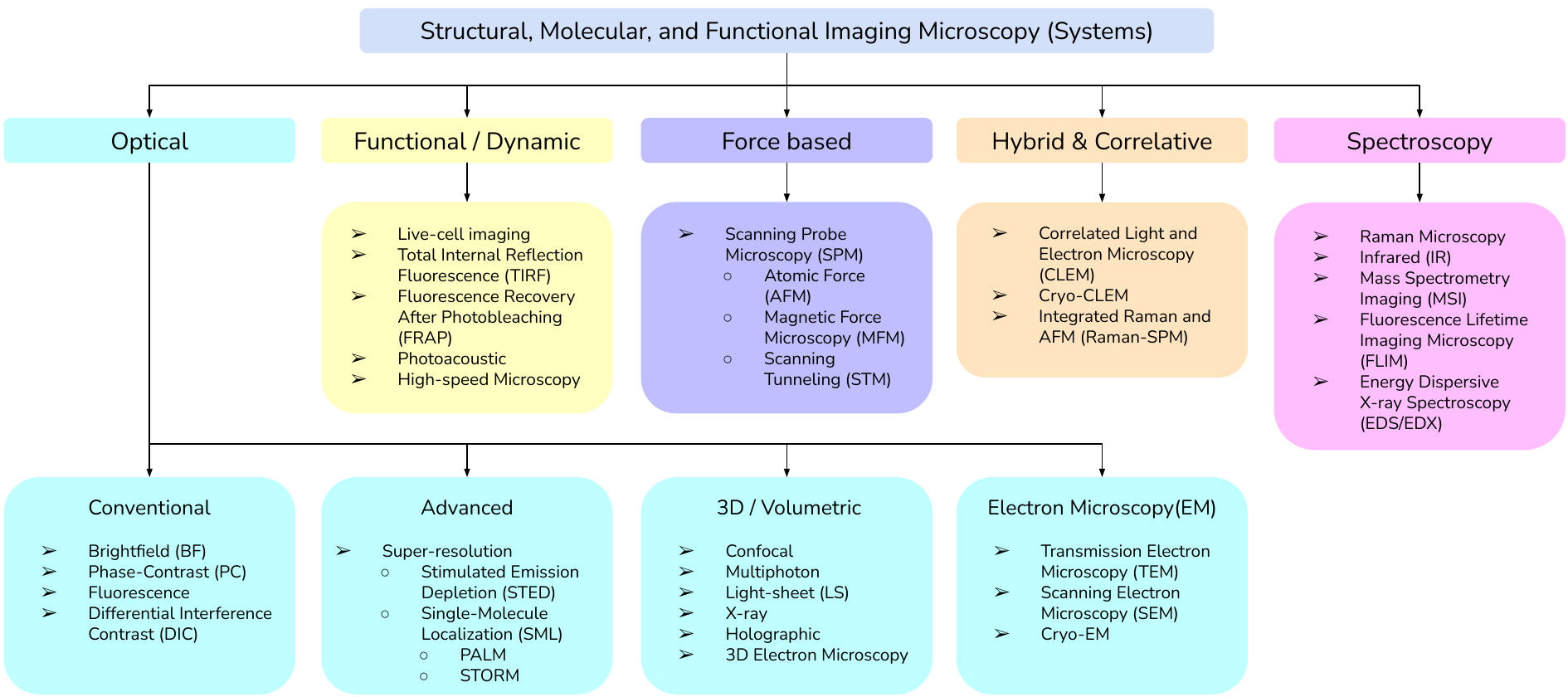}
\caption{A comprehensive classification of diverse imaging and analytical methods, organized into structural, molecular, functional, and hybrid approaches.}
\label{fig: microscopy stratification}
\end{figure}

\textbf{Types of image based microscopy.}
Fig.~\ref{fig: microscopy stratification} illustrates a comprehensive hierarchical classification of diverse imaging methods in microscopy. This classification is structured around three primary functional objectives: (1) visualizing fine structures through static imaging, (2) tracking dynamic processes such as motion and molecular interactions, and (3) probing molecular composition to reveal chemical or elemental properties. Techniques are further grouped based on their underlying principles, including spectroscopy-based, electron-based, force-based, and light-based approaches, with subdivision into conventional, advanced, and specialized types. This organization captures the broad range of tools available for structural visualization, molecular analysis, dynamic behavior tracking, and multimodal integration. Among the commonly used techniques are confocal microscopy, phase-contrast microscopy, fluorescence microscopy, quantitative phase microscopy, atomic force microscopy (AFM), and scanning electron microscopy (SEM). Each method has unique advantages and limitations. For instance, confocal microscopy enables optical sectioning for 3D imaging but requires longer acquisition times, whereas phase-contrast microscopy provides label-free visualization of live cells but lacks molecular specificity. While the classification provides clarity, the boundaries between categories are not absolute. Many methods exhibit properties that span multiple functions. For example, total internal reflection fluorescence (TIRF) microscopy is fluorescence-based and can be kept under optical microscopy, but it is classified under dynamic imaging due to its primary application in studying near-surface events. Also, this classification is not exhaustive. Numerous techniques and their variants, including emerging modalities and niche applications, remain outside the scope of this review. However, the techniques presented here represent some of the most widely used and impactful methods in life science research.

There are also other noteworthy binary classifications that offer valuable perspectives.
\textit{\underline{Fluorescence vs label-free}}: fluorescence microscopy employs chemical or genetic tags to label specific structures, whereas label-free methods (e.g. phase-contrast) do not require exogenous markers. In Fig.~\ref{subfig: microscopy image types}, for instance, fluorescence distinctly visualizes nuclear envelopes and nuclei, while phase-contrast records all features in a single image without labels from the same sample.
\textit{\underline{Transparent vs opaque samples}}: techniques based on light transmission (e.g. brightfield, phase-contrast, fluorescence) require minimal preparation and support live specimen imaging but are ineffective for highly scattering or opaque samples. In such cases, reflection or scattering methods (e.g. SEM, TEM, AFM, TIRF) are more suitable.
\textit{\underline{Thin vs thick specimens}}: techniques like TEM excel with thin samples, offering high resolution but limiting live or intact specimen studies. Confocal and multiphoton microscopy handle thick samples, preserving 3D structures but with reduced resolution in deeper layers.
\textit{\underline{Live-cell vs static samples}}: live-cell imaging captures dynamic processes but is limited by phototoxicity and lower resolution. Fixed-sample methods, such as electron microscopy, provide detailed static images but lack the ability to track ongoing biological events.
Overall, image-based microscopy is essential for studying biological structures at various scales but faces many challenges due to its complexity, as further discussed in the next section.

\section{Challenges in image based microscopy and AI}
Image based microscopy holds immense potential for transformative applications and advancements through the integration of AI. However, the field faces several critical challenges, including the lack of sufficient labeled data, variability in data quality, and the pervasive presence of noise and artifacts in microscopy datasets. In this section, we delve into some of the key obstacles in the field, highlighting their implications.

\textbf{Fluorescence vs label-free microscopy.}
In microscopy, both label-free and fluorescence techniques are vital for imaging biological processes, yet each has inherent trade-offs and challenges. Label-free methods, such as phase-contrast or quantitative phase microscopy, preserve the natural state and vitality of samples, making them ideal for live-cell imaging. However, they often lack molecular specificity i.e blurred edges, making it difficult to isolate or study particular biomolecules or pathways. Conversely, fluorescence microscopy offers high specificity and sensitivity, enabling the visualization of targeted molecules. Yet, it can disrupt the sample’s natural state, reduce viability through phototoxicity, and cause off-target binding, potentially compromising reproducibility \cite{labelfree2023}. These limitations highlight the complementary nature of the two approaches and the need for innovative methods to bridge the gap between specificity and sample preservation.

\textbf{Data labeling problem.}
Labeling biological data for AI applications, such as segmentation and tracking, is a significant challenge due to the complexity, size, and dynamic nature of biological systems. Biological entities often interact and exhibit intricate dynamics, requiring extensive and precise annotations to define distinct features for algorithm training \cite{digitalstain2019}. This task is labor intensive and prone to errors, as biological images often involve large datasets with subtle variations that are difficult to interpret consistently. The lack of sufficient labeled data hampers the ability to train and verify robust AI systems, creating a bottleneck for advancements in biological image analysis \cite{holostain2020,nmi2021dkp}. Addressing this issue requires innovative solutions, such as semi-supervised learning, active learning, or leveraging synthetic data to reduce the dependency on manual labeling.

\begin{figure}[t]
    \centering
    \begin{subfigure}[b]{0.37\textwidth}
        \centering
        \includegraphics[width=\textwidth]{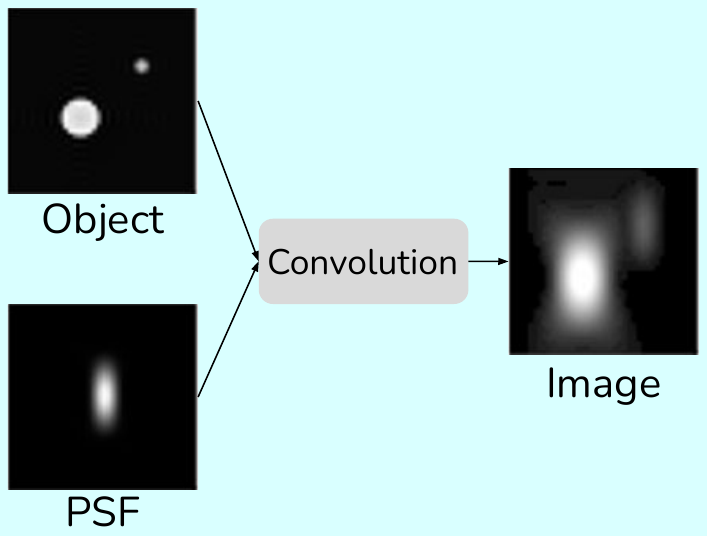}
        \caption{Image formation}
        \label{subfig: psf formation}
    \end{subfigure}
    \hfill
    \begin{subfigure}[b]{0.56\textwidth}
        \centering
        \includegraphics[width=\textwidth]{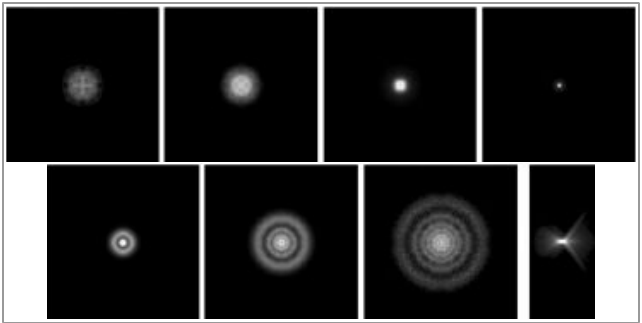}
        \caption{PSF types}
        \label{subfig: psf types}
    \end{subfigure}
    \caption{\textbf{Left}: PSF affecting the formation of an image by blurring the real object. \textbf{Right}: Different types of empirical PSFs in the green emission range, for a $\times 100$ 1.4 \textit{NA} objective, with an oil immersion refractive index of $n=1.518$ \cite{psftypes2005}.}
    \label{fig:model}
\end{figure}

\textbf{Point spread function (PSF).}
It is a fundamental characteristic of a microscopy system that quantifies how a point source of light is imaged and represented by the system. It describes the response of the optical setup to a point source, effectively encapsulating how the inherent physical and optical properties of the system distort or blur the representation of an object. Mathematically, the PSF represents the intensity distribution of light in the image plane resulting from a point source located at the focal plane \cite{psfdefinition2023}. Fig.~\ref{subfig: psf formation} demonstrates the formation of the output image as the object is convolved with the PSF, leading to image blurring. Fig.~\ref{subfig: psf types} shows different types of PSFs. To estimate the PSF, beads of known shapes are imaged under the microscope (see details in Sec.\ref{sec:synergetic relation}). Due to the diffraction limit and the nature of PSF, multiple distinct objects can produce similar output images, making the inverse problem ill-posed. This ambiguity presents significant challenges in training AI systems, even with labeled data, as the many-to-one mapping often leads to suboptimal convergence of AI training algorithms.

\textbf{Noise types and their impact.}
Various factors contribute to noise in microscopy images, including variation in PSFs, the process of image formation, optical aberrations, disturbances caused by mismatched refractive indices within the sample, out-of-focus sample and light originating from out-of-focus regions. Additionally, variations in sample preparation methods further exacerbate these challenges \cite{psftypes2005}. Noise characteristics can range from well defined types, such as textured noise, which can be modeled and estimated with known approaches \cite{sarcgraph2021plosone}, to more complex noise that resists straightforward characterization and modeling \cite{nmi2021dkp}. AI algorithms trained to handle specific types of noise often fail to generalize effectively, resulting in suboptimal performance when applied to images of the same biological sample but affected by different noise types.

\textbf{Dynamic nature of biological events.}
Modeling interactions among biological structures presents significant challenges due to their inherently dynamic behavior. These structures often exhibit continuous motion, frequently moving in and out of the imaging field of view. This characteristic of biological samples complicates their tracking and makes their interaction analysis highly challenging. Furthermore, the rapid and often non linear nature of these movements adds additional complexity to the extraction of meaningful insights \cite{celltrackingchallenge2017naturemethods}.

\textbf{Cellular vs subcellular level microscopy.}
Cellular level microscopy typically operates within the diffraction limit, providing clear boundaries and well defined structures. In contrast, subcellular imaging often exceeds the diffraction limit, leading to challenges in resolving fine structural details and demarcating boundaries. This limitation results in reduced resolution, complicating data analysis and hindering AI model training. Similar issues arise in nanoscale imaging of subcellular interactions, which is an active area of research. Advanced techniques, such as Single Molecule Localization Microscopy and correlative microscopy, have been developed to overcome these challenges and enhance resolution at the subcellular scale.

\textbf{Toxicity and bleaching.}
Phototoxicity and cytotoxicity pose significant challenges in live-cell imaging, as prolonged exposure to light or staining agents can compromise cellular viability and physiology, leading to artifacts. Additionally, fluorophore bleaching during imaging results in signal loss, reducing the quality and reliability of long term observations and quantitative analyses \cite{microscopytypes2022algorithms}.

\textbf{Multidisciplinary nature of research.}
Effective collaboration in this multidisciplinary field may be hindered by communication gaps, including technical jargon and unclear task delegation, such as deciding who should label data for preliminary analysis. Biologists typically have the domain expertise but limited time, whereas informaticians, while more available, might lack the necessary expertise. This dilemma can impede project efficiency and compromise data quality.

\section{Possible solutions}
Because these challenges are inherently multidisciplinary, strengthening collaboration among researchers from diverse fields is crucial for addressing them effectively. Collaboration fosters innovative problem solving and efficient sharing of resources. This section outlines potential solutions to the identified challenges.

\textbf{Synthetic data.}
Synthetic data addresses the challenge of limited labeled data by providing an alternative source for training and testing models. It can be generated through simulators \cite{nmi2021dkp,sarcgraph2021plosone}, which approximate real-world data generation processes, or via generative AI models \cite{emdatagen2024arxiv}, offering flexibility to create abundant datasets with distributions resembling real data.

\textbf{Physics-based AI.}
Physics-based AI helps in training AI models by reducing their dependence on labeled datasets by either incorporating the physics constraints during the training data generation process \cite{sarcgraph2021plosone}, or by modifying the design of an AI model to bypass its dependency on labeled data by incorporating physics-based priors into the model \cite{physicsbasedml2021,superres2021pssrsimulation}. Physics-based models not only overcome data scarcity but also often yield models with improved interpretability.

\textbf{Cross-microscope or cross-noise distribution models.}
Cross-modal (or cross-microscope) approaches can effectively address challenges such as data scarcity or the lack of labeled datasets. These methods leverage complementary modalities either for information fusion \cite{digitalstain2019} or by incorporating priors extracted from one modality into the analysis of another \cite{superres2021pssrsimulation,punnakkal2023mishape}. Such cross-modality strategies are particularly valuable in tackling subcellular imaging challenges, where the integration of information across modalities enhances resolution and interpretability, and reduces toxicity. Additionally, these challenges can be conceptualized as problems of distributional shifts, where domain adaptation techniques offer robust solutions to align disparate data distributions and mitigate noise \cite{domainadaptation2024imclw}.

\textbf{Self supervised learning (SSL).}
SSL is a machine learning paradigm that leverages large amounts of unlabeled data to derive meaningful representations through automatically generated supervisory signals. In microscopy, both live-cell and static imaging applications benefit from SSL by employing tailored objective functions, such as contrastive learning, to effectively extract biologically relevant features from unlabeled datasets \cite{sslimage2023miccai,sslsupercut2024microscopy,rizk2014segmentation}.

\textbf{Transfer learning.}
Because of large scale training on diverse datasets, pretrained models often capture robust and broadly applicable feature representations, surpassing those learned from scratch. Consequently, they can be efficiently finetuned for new tasks when domains overlap and labeled data are scarce \cite{transferlearning2021applsci}. Another progressive approach within transfer learning is \textit{active learning}, wherein each generation of models benefits from knowledge transferred by the previous one. This methodology has enabled the development of foundational models such as \cite{segmentanything2024medical}, which demonstrate high generalizability and can accelerate training in low-labeled-data contexts. Additionally, transfer learning facilitates multi-task learning across various imaging applications \cite{transfermultitask2021cbm}.

\section{Current and future areas of research}
In this section, we examine current and prospective research areas in life science. Each area includes its primary challenge, followed by secondary challenges, then its classification as either current or future, and key references. Current areas are those undergoing active investigation, whereas future areas are those where minimal or no research has yet been conducted.

\textbf{Video analysis.}
Involves investigation of microscopic videos for activity recognition, event detection, tracking, event detection in untrimmed videos, detection of dynamic events (e.g. entities moving in and out of the frame), and unsupervised event clustering \cite{eventdetection2024biorxiv}. \textit{\underline{Future}}.

\textbf{Amodal segmentation.}
Segmentation is an essential precursor to downstream tasks such as tracking or event analysis. Contemporary methods, including semantic or instance segmentation, often fail to handle overlapping structures adequately, resulting in incomplete analyses. These limitations can be addressed by developing amodal segmentation algorithms, which infer hidden object regions beyond visible boundaries. \textit{\underline{Future}}.

\textbf{Morphology analysis.}
Involves segmenting and quantifying shape and size of diverse biological structures. This remains especially demanding when conducted fully automatically. Subfields may include segmentation, morphological characterization, metric learning \cite{nmi2021dkp,punnakkal2023mishape}. \textit{\underline{Current}}.

\textbf{Anomaly detection.}
Involves grouping or modeling irregular behaviors to identify anomalous events that may serve as early disease indicators. Subfields include clustering, anomalous event analysis, and morphological anomaly detection \cite{anomlaydet2022}. \textit{\underline{Future}}.

\textbf{Application of foundation models.}
Involves leveraging large scale foundation models (e.g. LLMs, SAM variants) for captioning, segmentation, and tracking in microscopy, using minimal labeled data. Subfields include refining transfer learning strategies for diverse imaging applications; language based annotation of events and morphology, as well as specialized segmentation frameworks \cite{segmentanything2024medical}. \textit{\underline{Current}}.

\textbf{Sustainable and efficient AI for microscopy.}
Microscopy data requires massive storage and computational power for processing, necessitating the development of energy efficient algorithms. Subfields include data compression, efficient model design, and lightweight models tailored for large scale image analysis; leading to scalable and environmentally responsible solutions \cite{lightweight2024segmentation}. \textit{\underline{Future}}.

\textbf{Event grounding.}
Involves prompt based (text or visual) retrieval of information within image or video data. Subfields may include target event grounding, biological entity grounding, symptom grounding (to facilitate confirmatory or risk factor analyses), specific interaction searches. \textit{\underline{Future}}.

\section{Resources}
Open science initiatives have significantly increased the availability of data repositories and tools for researchers in this field. Prominent research institutions include the European Bioinformatics Institute (EBI), the National Cancer Institute (NCI), the European Molecular Biology Laboratory (EMBL), the Janelia Research Campus (JRC), the RIKEN Center for Biosystems Dynamics Research (BDR), and the MRC Laboratory of Molecular Biology (LMB). Below is a list of useful data repositories and tools.

\begin{itemize}
    \item BioImage Archive: \href{https://www.ebi.ac.uk/bioimage-archive}{https://www.ebi.ac.uk/bioimage-archive}
    
    \item Dataverse (hosted by different countries): \href{https://dataverse.org/installations}{https://dataverse.org/installations}
    
    \item Electron Microscopy Data Bank (EMDB): \href{https://www.ebi.ac.uk/emdb}{https://www.ebi.ac.uk/emdb}

    \item The Cancer Imaging Archive (TCIA) \href{https://www.cancerimagingarchive.net}{https://www.cancerimagingarchive.net}

    \item SciLifeLab: \href{https://www.scilifelab.se/data/repository}{https://www.scilifelab.se/data/repository}

    \item The cell: \href{https://www.cellimagelibrary.org/pages/datasets}{https://www.cellimagelibrary.org/pages/datasets}

    \item Image Data Resource (IDR): \href{https://idr.openmicroscopy.org}{https://idr.openmicroscopy.org}

    \item Zenodo: \href{https://zenodo.org}{https://zenodo.org}

    \item NeuroVault: \href{https://neurovault.org}{https://neurovault.org}

    \item BrainMaps: \href{https://brainmaps.org}{https://brainmaps.org}
\end{itemize}

\begin{itemize}
    \item DeepCell \href{https://www.deepcell.org}{https://www.deepcell.org}

    \item PyTorch: \href{https://pytorch.org}{https://pytorch.org}

    \item \href{https://imagej.net/ij}{https://imagej.net/ij}

    \item Royal Microscopical Society (RMS): \href{https://www.rms.org.uk/library/online-microscopy-resources/image-analysis-resources.html}{https://www.rms.org.uk}

    \item MicroscopyDB: \href{https://microscopydb.io}{https://microscopydb.io}

    \item ZEISS: \href{https://zeiss-campus.magnet.fsu.edu/index.html}{https://zeiss-campus.magnet.fsu.edu/index.html}
\end{itemize}

\section{Conclusion}
This work provides a comprehensive introduction to microscopy and its relationship to both life science and AI. A detailed classification of different microscopy techniques, alongside key research challenges, is offered to establish a foundational and experimental understanding of the imaging process. We have also explored various obstacles and proposed potential solutions, while highlighting a number of future application ideas and relevant resources. It is our hope that this work equips researchers with the necessary background, insights, and tools to accelerate their efforts.



\end{document}